\documentclass[acmtog, screen, nonacm]{acmart}

\AtBeginDocument{%
  }

\setcopyright{rightsretained}
\copyrightyear{2024}
\acmYear{2024}
\acmDOI{XXXXXXX.XXXXXXX}

\usepackage{booktabs} 
\usepackage{bm}
\usepackage{wrapfig}

\newlength\savedwidth
\newcommand\whline[1]{\noalign{\global\savedwidth\arrayrulewidth
                               \global\arrayrulewidth #1} %
                      \hline
                      \noalign{\global\arrayrulewidth\savedwidth}}

\begin{document}

\title{Elastic Locomotion with Mixed Second-order Differentiation}

\author{Siyuan Shen}
\email{messyshen@gmail.com}
\author{Tianjia Shao}
\email{tjshao@zju.edu.cn}
\author{Kun Zhou}
\email{kunzhou@acm.org}
\affiliation{%
  \institution{State Key Laboratory of CAD\&CG, Zhejiang University}
  \city{Hangzhou}
  \country{China}
}

\author{Chenfanfu Jiang}
\affiliation{%
  \institution{UCLA}
  \city{Los Angeles}
  \country{USA}}
\email{chenfanfu.jiang@gmail.com}

\author{Sheldon Andrews}
\affiliation{%
  \institution{École de technologie supérieure (ÉTS)}
  \city{Montreal}
  \country{Canada}
  }
\affiliation{
  \institution{Roblox}
  \city{San Mateo}
  \country{USA}
}
\email{sheldon.andrews@etsmtl.net}

\author{Victor Zordan}
\affiliation{%
  \institution{Roblox}
  \city{San Mateo}
  \country{USA}}
\email{vbzordan@roblox.com}

\author{Yin Yang}
\authornote{Corresponding author.}
\affiliation{%
  \institution{University of Utah}
  \city{Salt Lake City}
  \country{USA}}
\email{yangzzzy@gmail.com}

\renewcommand{\shortauthors}{Shen et al.}


\begin{abstract}
We present a framework of elastic locomotion, which allows users to enliven an elastic body to produce interesting locomotion by prescribing its high-level kinematics. We formulate this problem as an inverse simulation problem and seek the optimal muscle activations to drive the body to complete the desired actions. We employ the interior-point method to model wide-area contacts between the body and the environment with logarithmic barrier penalties. The core of our framework is a mixed second-order differentiation algorithm. By combining both analytic differentiation and numerical differentiation modalities, a general-purpose second-order differentiation scheme is made possible. Specifically, we augment complex-step finite difference (CSFD) with reverse automatic differentiation (AD). We treat AD as a generic function, mapping a computing procedure to its derivative w.r.t. output loss, and promote CSFD along the AD computation. To this end, we carefully implement all the arithmetics used in elastic locomotion, from elementary functions to linear algebra and matrix operation for CSFD promotion. With this novel differentiation tool, elastic locomotion can directly exploit Newton's method and use its strong second-order convergence to find the needed activations at muscle fibers. This is not possible with existing first-order inverse or differentiable simulation techniques. We showcase a wide range of interesting locomotions of soft bodies and creatures to validate our method. 
\end{abstract}

\begin{teaserfigure}
\center
  \includegraphics[width=\textwidth]{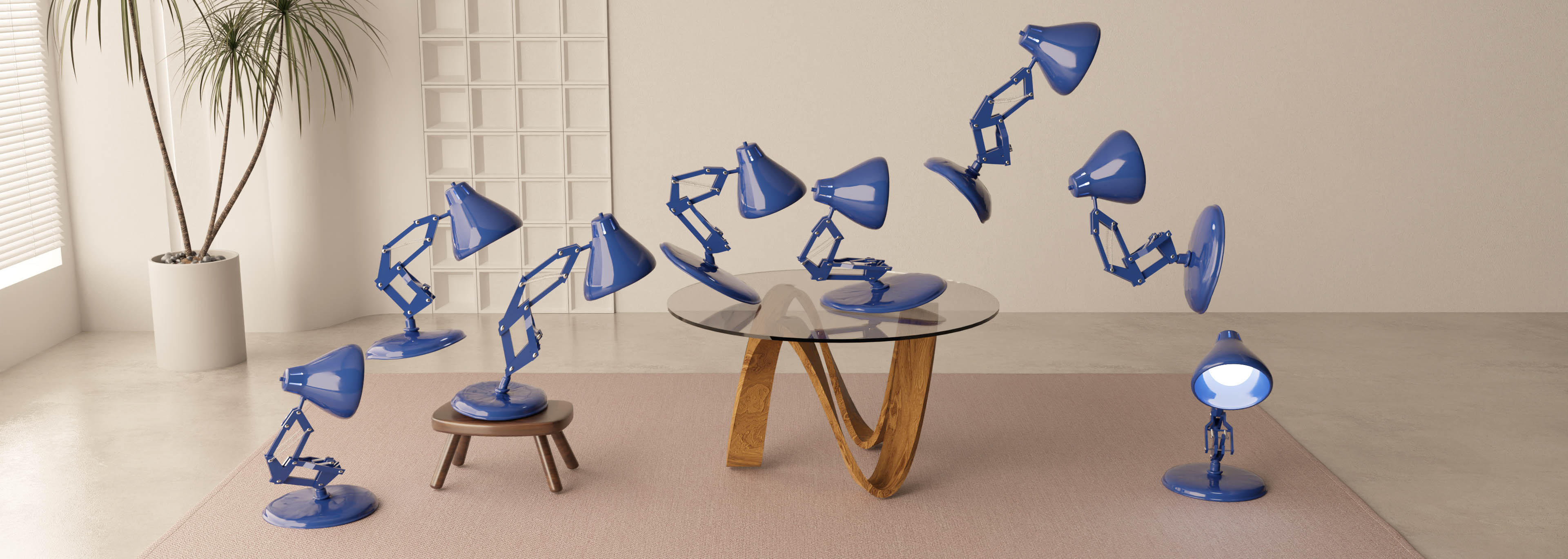}
  \caption{\textbf{A gymnast lamp.}~~Elastic locomotion offers the user a convenient way to bring elastic bodies to life. By specifying intuitive and high-level motion descriptions, our framework finds the optimal muscle activations that drive the body to achieve those kinematic targets. The key to solve this problem is a mixed second-order differentiation that augments reverse automatic differentiation with complex-step finite differences. This allows the Hessian of a user-defined loss function to be conveniently evaluated. In this teaser example, a lamp first jumps onto a small stool, then onto a higher glass table (they have different surface frictions), performs a flip and twist in mid-air, and finally lands on the ground. It takes about $20$ sec to optimize each time step ($\Delta t = 1/40$ sec), and we only need two Newton iterations on average to solve the inverse problem. This is not possible with existing first-order methods.}
  \label{fig:teaser}
\end{teaserfigure}


\maketitle

\section{Introduction}\label{sec:intro}
The animation of deformable characters is a desirable goal for bringing real-world and fantastic characters to life. While many characters rely on an underlying piece-wise rigid skeleton rig to achieve coordinated motions, there are also a lot of creatures that do not have embedded bones -- they exploit muscular actuators to contract or expand parts of the body to enable unique and vibrant interactions with the environment. 
Unlike a skeleton rig, a soft body has a flexible and extensive surface, leading to wide contacting areas with the environment e.g., the ground. As a result, calculating a good locomotion control of the soft body inevitably needs to deal with a high-dimension contact problem involving a set of inequality constraints. While the mathematical model of contact has been well-studied, e.g., one can simplify it to linear complementarity programming~\cite{cottle2009linear}, it is known NP-Hard~\cite{chung1989np} due to its combinational nature and remains computationally challenging in practice. 

Most real-world materials are nonlinear. Forward simulation of a soft body by specifying external forces is already expensive as tens of thousands of unknown degrees of freedom (DOFs) are two-way coupled. It becomes tricky if the user wants direct and intuitive control over the result of the simulation, e.g., to have movement of the body to follow a prescribed path. This is because the objective function and the control parameters are implicitly related by the forward process without a closed-form mathematical expression, and this implicit relation also involves a set of contact constraints. Such combined high nonlinearity and high dimensionality are beyond the capability of the existing simulation techniques, making the problem of elastic locomotion particularly challenging. 

We develop a framework that enables an intuitive control of an elastic body's locomotion. The user directs the motion, pose, or trajectory of the body via high-level and straightforward kinetic descriptors, and our algorithm inversely estimates the optimal muscle activations to achieve the desired dynamic state at the next frame. We are inspired by incremental potential contact or IPC~\cite{liIncrementalPotentialContact2020a}. IPC re-formulates inequality constraints using logarithmic barrier penalties and converts the contact problem to an unconstrained optimization. This strategy allows us to tackle higher-dimensional contact instances since IPC obviates the combinational optimization inherent in the complementarity programming. On the downside, IPC further escalates the nonlinearity of the forward simulation. The resulting optimization is bumpy and irregular and has a large number of saddle points and local minima. Existing differentiable or inverse simulation techniques, which are primarily based on gradient information of the loss function are barely helpful. 

This paper proposes a novel mixed differentiation modality that allows an efficient evaluation of second-order differentiation for elastic locomotion so that we can use high-order optimization tools like Newton's method to overcome the nonlinearity induced by inverse elastodynamics and IPC barriers. The core idea is to combine analytic differentiation of auto differentiation (AD)~\cite{bartholomew2000automatic,corliss2013automatic}, which generates a programming procedure of the analytic differentiation of a function, with numerical differentiation, which estimates the differentiation of a function without knowing its closed-form. Specifically, we employ complex-step finite difference (CSFD)~\cite{luo2019accelerated,shenHighorderDifferentiableAutoencoder2021} as our numerical differentiation method and augment CSFD with reverse AD. Compared with other numerical differentiation methods such as finite difference, CSFD has several favored properties. First, CSFD is numerically stable and accurate. It does not have the subtractive cancellation issue, so one can choose a sufficiently small perturbation size to fully suppress the approximation error, making CSFD as accurate as the analytic result. There are many off-the-shelf complex libraries. Thus, we do not need to build our differentiation method from scratch.


\section{Related Work}\label{sec:related}
In this section, we briefly discuss some prior contributions relevant to soft body control, trajectory optimization, and differentiation. 
\subsection{Soft body Control}
The appeal of animated soft body controllers has led to the introduction a number of techniques offering solutions even with the significant challenges of slow simulation and of complexity in the specification of muscles and/or activation. Much of the early work in this area
leaned heavily on characters with simple complexity to address speed~\cite{tan2012soft,coros2012deformable,kim2011fast} while each utilitize different mechanics for activation.  Namely, \citet{kim2011fast}
embed a skeleton into deformable body to drive animation, while \citet{coros2012deformable}
build controllers from the soft-body posed into target
states.  Our muscle system more closely matches that of~\cite{tan2012soft}, as we model muscles as spring-like curves that add forces along their attachments in their current direction.  
\citet{barbivc2009deformable} employs a reduced order model 
to speed up their deformation simulation, along with target shapes created manually.  Their pioneering work animates soft-body characters through the reduced simulation with the controller running in the low dimensional subspace of the reduced model~\cite{barbivc2008real,barbivc2009deformable}.

While a tremendous amount of work has been introduced in the last decade that employs
deep reinforcement learning (DRL) in the area of articulated control~\cite{DRLsurvey},
we have seen limited work that employs DRL for soft-body control.  One exception to this is SoftCon~\cite{min2019softcon}, a deformable simulation extension of~\cite{Tan:2011a} for animating underwater creatures with several advances.  \citet{min2019softcon} combines a novel muscle excitation/ propagation model with DRL to successfully learn the control policies of soft-bodied animals.  Their 
work employs FEM~\cite{sifakis2012fem}
over triangulation~\cite{tetwild} derived for relatively low resolution character models.  Nonetheless, their characters are among the most expressive of those generated showcasing a wide range of behavior results.
There is also increasing interest in the development of DRL gym frameworks for soft body problems~\cite{schegg2022sofagym}.

It is also worth noting that parallel work has been ongoing in soft robotics~\cite{Cheney2014,Coevoet2017, Della2023}, with several 
cross-over works, among these are~\cite{liang2023learningreducedordersoftrobotcontroller,bern2019trajectory,hu2019chainqueen}.  


\subsection{Trajectory Optimization}

Given keyframe information from the user, the
production of control using out differential solver bares some resemblance to work in trajectory optimization.  Rather than
cast the simulation control as a initial value problem with iterative search, trajectory optimization seeks to solve an entire trajectory
simultaneously.  Early work in this area, so-called spacetime solvers cast
very simple dynamics as constraints that were upheld as acceleration values were computed~\cite{witkinkass,cohen1992interactive,ngo1993spacetime}.

Since the introduction of spacetime optimization, several researchers have proposed mechanisms for incorporating trajectory optimization to solve varying character animation control problems~\cite{liu2005learning,mordatch2014combining,safonova2004synthesizing,bern2019trajectory,zordan2014control,al2012trajectory}, among others.
A key difference with our problem domain is the incorporation of soft-body dynamics.  Deformable simulation introduces significantly more complex contact conditions, which foils many of the undelying assumptions of the previous work.  As such, to solve our desired control problem, we introduce a novel differentiation tool that allows efficient second-order evaluation so that we can use high-order optimization to overcome the inherent nonlinearities. 

\subsection{Differentiation \& differentiable simulation}
Evaluating the derivative for differentiation of the loss function is an indispensable step for inverse problems. A commonly used differentiation technique is automatic differentiation or AD~\cite{griewank2008evaluating,bartholomew2000automatic,corliss2013automatic}, which decomposes complicated functions with the chain rule. AD has been extensively used in graphics~\cite{grinspun2003discrete,guenter2007efficient}. Back propagation~\cite{hecht1992theory} widely used by the machine learning community 
training is a special implementation of the reverse AD. AD also forms the foundation of many differentiable simulation frameworks~\cite{du2021diffpd,hu2019difftaichi,hu2019chainqueen}. While one may perform AD multiple times to obtain a high-order derivative, it has been argued that recursively applying AD leads to inefficient and numerically unstable code~\cite{betancourt2018geometric,margossian2019review}. 

In contrast to AD, numerical differentiation is also popular. The standard approach is the finite difference method, which however suffers from the \emph{subtractive cancellation} issue: decreasing
the magnitude of the perturbation will diverge the calculation~\cite{brezillon1981numerical}. CSFD is a finite difference scheme based on the complex version of Taylor series expansion~\cite{lyness1967numerical}. It has been used for nonlinear FEM~\cite{luo2019accelerated,kim2011numerically} and sensitivity analysis~\cite{montoya2015finite,anderson2001sensitivity}. While it is possible to directly use CSFD to compute high-order derivatives~\cite{shenHighorderDifferentiableAutoencoder2021} by generalizing regular CSFD to multi-complex numbers. It requires excessive forward function evaluations and is prohibitive if the actual derivative is needed (e.g., in Newton's method). 

We posit that achieving high-order differentiation for general-purpose and complicated computations should be enabled in a hybrid way, by leveraging the best of both numerical and analytic differentiation modalities. This allows an efficient gradient calculation without significantly increasing the order of the differentiation in each modality. To this end, we propose a novel CSFD-AD differentiation scheme. By treating AD as a black-box computation and promoting it with CSFD, we can conveniently obtain the gradient of AD's output or, subsequently, the Hessian of the target loss.  

Our method is also relevant to differentiable IPC simulation~\cite{huang2022differentiable}, which uses the adjoint method~\cite{givoli2021tutorial} to calculate the loss gradient. Unfortunately, high-order adjoint method remains a research problem, and it becomes less help for elastic locomotion, for which second-order descent information is essential.  Our method, on the other hand, can effectively obtain the Hessian for the use of higher-order optimizations to optimize the target loss, which is very useful for dealing with complex collision situations.





\begin{figure*}
  \centering
  \includegraphics[width=\textwidth]{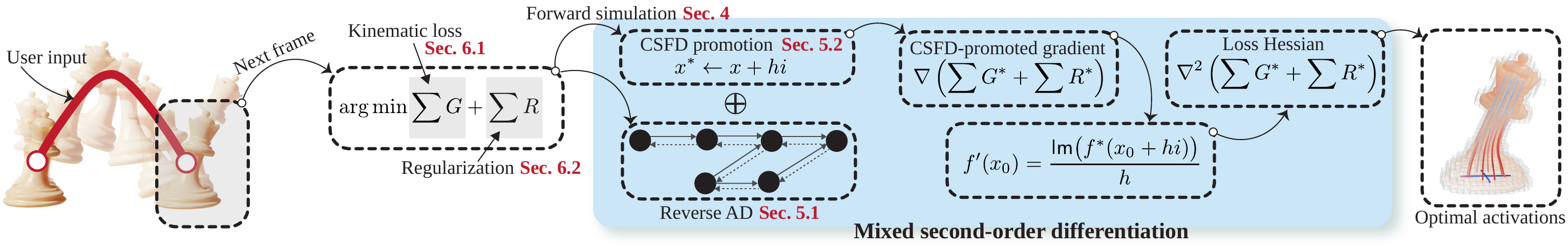}
  \caption{\textbf{An overview of our computational pipeline.}~~Elastic locomotion solves an inverse simulation problem. The user specifies high-level kinematic targets the body expects to achieve, which is formulated as a scalar-value loss function combining the target function and regularization terms. We use classic line-search Newton to solve this problem. The Hessian of the loss function is obtained by our mixed second-order differentiation scheme, which combines CSFD and AD to efficiently calculate contact-in-the-loop loss Hessian.}\label{fig:pipeline}
\end{figure*}
\section{Overview}\label{sec:overview}
Fig.~\ref{fig:pipeline} outlines major steps of our pipeline and their corresponding discussions in the paper. Elastic locomotion embodies a standard inverse problem. Given a soft object embedded in a set of both contractible and extensible muscle fibers, elastic locomotion estimates necessary activation forces along the muscle fibers that enable natural and coordinated movement of the object under user-prescribed high-level motion descriptors e.g., the body trajectory. While muscle activations are the driving force, it is frictional contacts between the environment that play a key role in the body's locomotion. Elastic locomotion decomposes the high-dimension space-time optimization problem into sequential nonlinear programming frame by frame. At each frame, we use a mixed differentiation modality, which efficiently returns the first- and second-order gradient information (w.r.t. the target loss function), and it is agnostic to the specific simulation algorithm adopted nor the total number of interactions used. The second-order descent direction is indispensable for such a strongly nonlinear inverse problem. In the following sections, we elaborate on each of the major steps in detail.

\section{Problem formulation}\label{sec:problem}
We discretize a 3D soft object with a tetrahedral mesh of $N$ nodes and embed muscle fibers in its body as the primary motion driver. The equation of motion of this deformable object is modeled using nonlinear FEM and implicit Euler, in the form of force equilibrium as: 
\begin{equation}\label{eq:equation_of_motion}
\begin{split}
    \bm{x}_{t+1} &= \bm{x}_t + \Delta t\bm{v}_{t+1}, \\
    \bm{v}_{t+1} &= \bm{v}_t + \Delta t\bm{M}^{-1}(\bm{f}_{int} + \bm{f}_{ext} + \bm{f}_d + \bm{f}_c + \bm{f}_m),
\end{split}
\end{equation}
where $\bm{M} \in\mathbb{R}^{3N \times 3N}$ is the mass matrix, $\bm{f}_{int}$, $\bm{f}_{ext}, \bm{f}_d$, $\bm{f}_c$, $\bm{f}_m$ represent elastic (i.e., internal), external, damping, contact and muscle forces, respectively, $\Delta t$ is the time step size, and $\bm{x}_t$ and velocities $\bm{v}_t$ at time step $t$.

The elastic force is the resultant force produced by the elastic properties of a material that resists deformation. It is calculated as the negative gradient of the elastic potential energy. In our implementation, we use the stable Neo-Hookean energy~\cite{smith2018stable}:
\begin{equation}
    \Psi = \frac{\mu}{2}(I_C - 3) - \mu(J-1) + \frac{\lambda}{2}(J-1)^2,
\end{equation}
where $I_C = \mathsf{tr}(\bm{F}^\top\bm{F})$ and $J = \mathsf{det}(\bm{F})$ are computed based on the deformation gradient $\bm{F}$. Terms $\mu$ and $\lambda$ are Lam\'{e} constants. Nevertheless, our pipeline is compatible with any hyperelastic material models. The external force, attributable to gravity, is constant. We use the Rayleigh damping to calculate the damping force:
\begin{equation}
    \bm{f}_d = \left(\alpha\bm{M} + \beta\bm{K}_0\right)\bm{v}_{t+1},
\end{equation}
where $\bm{K}_0$ is the rest-shape stiffness matrix. The contact force is evaluated following the IPC formulation. IPC activates a barrier potential energy $B$ if a surface triangle is sufficiently close to a collider:
\begin{equation}\label{eq:barrier}
    B(d, \hat{d}) = \left\{
    \begin{array}{ll}
    \displaystyle -\kappa \big(d - \hat{d}\big)^2 \ln \left(\frac{d}{\hat{d}}\right), & 0 < d < \hat{d},\\
    \displaystyle 0,& d \geq \hat{d}.
    \end{array}
    \right.
\end{equation}
Here, $d$ stands for the distance between the triangle and the collider; $\hat{d}$ is the constraint tolerance ($\hat{d} = 1E-3$ in our implementation); and $\kappa$ is a hyper-parameter controlling the initial stiffness of the barrier. The negative gradient of the barrier gives the collision force at (nearly) contacting triangles. The friction force is evaluated in a lagged way based on the information of the collision force in the previous iteration. 


Similar to~\cite{tan2012soft}, we model muscle fibers as polygonal curves with $M$ segments. Each muscle segment is modeled as an individual spring that can contract or extend along its current direction, but it does not bend. A segment exerts an activation $a \in \mathbb{R}$ along its direction, either contraction or extension. 
An activation affects multiple nearby elements. For the $i$-th element and  $j$-th muscle segment, the influencing weight $w_{ij}$ is based on their geodesic distance $g_{ij}$ on the mesh as a Gaussian kernel:
\begin{equation}
    w_{ij}=\mathsf{exp}\left(-\frac{g_{ij}^2}{c^2}\right),
\end{equation}
where $c$ is the variance of the Gaussian function. $w_{ij}$ only depends on the rest shape of the body and muscle, and it can be pre-computed. 
The accumulated muscle stress in the deformed coordinates of the element is:
\begin{equation}
\sigma_i=\sum_j w_{ij} \bm{R}\bm{E}_j\bm{R}^\top,
\end{equation}
where 
\begin{equation}
\bm{E}_j=\bm{U}
\begin{bmatrix}
f_j & 0 & 0 \\
0 & 0 & 0 \\
0 & 0 & 0 \\
\end{bmatrix}
\bm{U}^\top = \bm{d}_j \otimes \bm{d}_j a_j.
\end{equation}
\setlength{\columnsep}{5 pt}
\begin{wrapfigure}{r}{0.55\linewidth}
    \includegraphics[width=\linewidth]{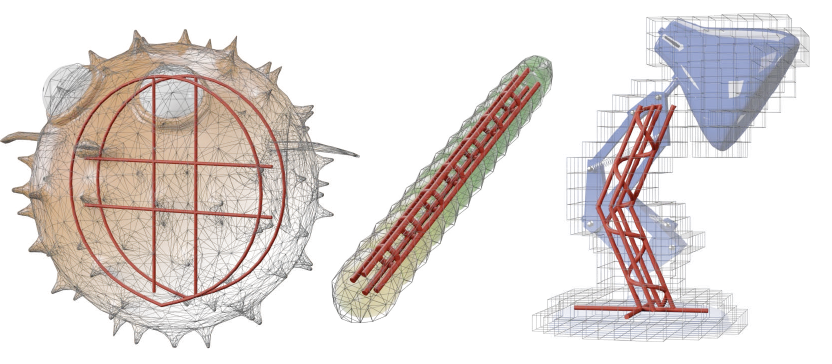}
    \caption{\textbf{Muscle arrangement.}~~We highlight the arrangement of muscles in some examples reported in the paper. We assume the muscle setup is provided.}\label{fig:muscle}
\end{wrapfigure}
Here, $\bm{R}$ transforms the stress $\bm{E}_j$ from the reference coordinates to the deformed coordinates, and $\bm{U}$ rotates vector $[1, 0, 0]^\top$ in the reference coordinates to align with $\bm{d}_j$, the direction of muscle segments in the reference coordinates.
We project the $\sigma_i$ to the area-weight face normal of the element in the deformed coordinates. This surface force is evenly split to vertices to derive $\bm{f}_m$ at every node. We use a pose-dependent activation matrix $\bm{A} \in \mathbb{R}^{3N \times M}$ to encode the muscle force computation, such that:
\begin{equation}
    \bm{f}_m = \bm{A}(\bm{x}) \bm{a} \,,
\end{equation}
where $\bm{a} \in \mathbb{R}^M$ is the vector of activations of all the muscle segments. 


Users can design the muscle arrangement in the deformable body to support different control tasks. Several muscle arrangements are shown in Fig.~\ref{fig:muscle}. For example, implanting several longitudinal muscle fibers spanning the height allows the body to be shortened through muscle contraction. Body bending can be manipulated via asymmetrical contraction, and body extension/stretching follows relaxing post-contraction. Circular muscles encircle the body, and radial muscles span a cross-section of the body. When these muscles contract, the body thins and gets lengthened due to the volume preserving of the material. Oblique muscles or helical muscles wrap around the body in a helical shape to support twisting. How to deploy such actuators for motion control itself is an interesting research problem~\cite{tzoumas2015minimal,thelen2003generating}. This however, is not the focus of this paper, and our system assumes a prescribed muscle placement.

\subsection{Loss function}
We express the objective function in the optimization as:
\begin{equation}\label{eq:loss}
    \arg \min_{{\bm{a}}} L(\bm{x}, \bm{a}) =  \sum \limits _{i}w_i G_i(\bm{x}_{t+1}, \bm{a}) + \sum \limits _{j} \lambda_j R_j(\bm{x}_{t+1}, \bm{a}),
\end{equation}
where each $G_i$ is a user-defined objective to depict the kinematic state of soft-body motion, such as jumping, twisting, and rolling. Regularization penalty terms $R_j$ are also included (Sec.~\ref{sec:regularization}). Our method solves for the appropriate muscle contraction and relaxation $\bm{a}$ to apply the muscle force $\bm{f}_m$, ensuring that the position $\bm{x}_{t+1}$ of the object at the next time step conforms to the user's description.


This is a challenging optimization problem because $\bm{x}_{t+1}$ depends on $\bm{a}$ implicitly via the equation of motion Eq.~\eqref{eq:equation_of_motion}, which needs to take into account dynamic contacts and friction. While the use of the barrier energy from Eq.~\ref{eq:barrier} obviates the need for complementarity programming, it increases the nonlinearity of the problem. Existing differentiable simulation methods~\cite{huang2022differentiable,du2021diffpd} are mostly first-order based on the gradient information of the loss function and fail to converge in our problem. 

Instead of relying on gradient exclusively, we leverage Newton's method to tackle the nonlinearity induced by contact barriers. Newton's method exhibits superior second-order performance in proximity of a local minimum. Before Newton iterations, we apply several iterations of gradient descent. Gradient descent is less sensitive to initial selection and offers a reasonable starting point for the follow-up Newton. Naturally, the muscle contraction of the last frame $\bm{a}_{t}$ serves as an initial guess in our process.





\section{Mixed Differentiation}\label{sec:mixdiff}
The method used to solve Eq.~\eqref{eq:loss} is far from novel -- it embodies a standard nonlinear procedure with line search. Yet, the challenge is not the optimization itself but the evaluation of the loss Hessian. 
\subsection{Reverse AD}
The loss function $L: \mathbb{R}^M \rightarrow \mathbb{R}$ maps an input activation vector to a scalar-value loss. A common practice is to use reverse AD to calculate its gradient, i.e., by back-propagating the differential from the output to the input, as used in most deep learning frameworks~\cite{hecht1992theory}\footnote{If the function maps a low-dimension input to a high-dimension output, forward AD is more efficient.}. 

In reverse AD, the computational process is performed in two phases: a forward pass and a reverse pass. The forward pass computes and stores function values and intermediate results. In the reverse pass, it passes derivatives backward from the output to the inputs, complementing each intermediate variable $v_i$ with an adjoint $\overline{v}_i=\frac{\partial y_i}{\partial v_i}$ representing the sensitivity of output $y_i$ w.r.t. changes in $v_i$. Take an intermediate step of $v_t = f_k(v_r, v_s)$ for example. Assuming $v_a$ and $v_b$ are not used in other intermediate computations, reverse AD applies the chain rule and gives $\overline{v}_a$ and $\overline{v}_b$ by:
\begin{equation}
    \overline{v}_r=\frac{\partial f_k}{\partial v_r}\overline{v}_t, \qquad \overline{v}_s=\frac{\partial f_k}{\partial v_s}\overline{v}_t.
\end{equation}

It is possible to solely use reverse AD to compute the Hessian of a function. Doing so involves two passes of reverse AD. The computational graph, as well as all the intermediate values generated during the first AD invocation, must be saved. The second reverse AD is applied at each component of the resulting gradient vector. This results in significant memory usage because each gradient component also generates a more expansive computational graph. In our problem, $\bm{x}_{t+1}$ is related to $\bm{a}$ via a barrier-in-the-loop FEM procedure, the memory consumption using AD for Hessian evaluation is prohibitive. In addition, repetitively applying AD for calculating high-order differentiation has been known to be numerically unstable~\cite{margossian2019review}. 

\subsection{CSFD}
Alternatively, we inject a different differentiation modality i.e., complex-step finite difference or CSFD, into this procedure to calculate the second-order differentiation of the loss function and avoid aforementioned issues due to consecutive AD invocations. Given a real-value function $f: \mathbb{R} \rightarrow \mathbb{R}$, assume that it is differentiable around $x = x_0$. Conventional finite difference scheme applies a small real perturbation $h$ to $x_0$ and approximates the function gradient as:
\begin{equation}
    f'(x_0) = \frac{f(x_0 + h) - f(x_0) }{h} + \bm{O}(h) \,.
\end{equation}
When $h$ gets smaller, $f(x_0 + h)$ and $f(x_0)$ become nearly equal to each other. Subtraction between them eliminates leading significant digits, and the result after rounding could largely deviate from the actual value. This stability issue is known as the subtraction cancellation. Therefore, finite difference is not suitable for our problem. 

CSFD promotes this function to the complex domain as $f^*:\mathbb{C}\rightarrow\mathbb{C}$. Here, we use $(\cdot)^*$ to denote a complex variable. CSFD applies the perturbation along the imaginary direction to estimate the gradient at $x^* = x_0 + 0i$ as:
\begin{equation}\label{eq:csfd}
f'(x_0)=\frac{\mathsf{Im}\big(f^*(x_0+hi)\big)}{h}+\bm{O}(h^2)\approx\frac{\mathsf{Im}\big(f^*(x_0+hi)\big)}{h}.
\end{equation}
Eq.~\eqref{eq:csfd} does not have a subtractive numerator, meaning it only has round-off error, regardless of the size of the perturbation $h$. This allows us to set $h$ sufficiently small to accurately estimate the gradient information. For instance, if $h \sim \sqrt{\epsilon}$ i.e., around $1\times10^{-16}$, CSFD approximation error is at the order of the machine epsilon $\epsilon$. Hence, CSFD can be as accurate as analytic derivative because the analytic derivative also has a round-off error of $\epsilon$. 

\subsection{Mixed differentiation}
CSFD can be generalized with multi-complex numerics to estimate the high-order differentiation~\cite{luo2019accelerated}. However, doing so needs to apply $\bm{O}(M^2)$ perturbations to the loss function to obtain each element in Hessian. This is prohibitive for a high-dimension inverse problem like ours. Unfortunately, in order to use Newton's method, the explicit Hessian matrix has to be assembled. To this end, we combine CSFD and AD to avoid excessive perturbations and an over-expansive computational graph. We treat AD as a generic function which maps the input $x$ to its first-order derivative. Applying CSFD of AD naturally gives us the second-order differentiation of the target function, and we name this mixed differentiation scheme CSFD-AD i.e., CSFD-perturbed (inverse) AD procedure. CSFD-AD follows the AD algorithm but promotes each intermediate variable and adjoint to complex variables: $v_i^*, \bar{v}_i^* \in \mathbb{C}$.

Second-order derivative of the $k$-th component of the input $[\nabla ^2 f]_k = \frac{\partial ^2 y}{\partial x_k^2}$ can be computed with a single application of mixed differentiation as: 
\begin{equation}\label{eq:csfd-ad}
    [\nabla ^2 f]_k =  \lim_{h\rightarrow 0}{\frac{\nabla f({\bm{x}+h\bm{e}_k})-\nabla f({\bm{x}})}{h}} \approx \frac{\mathtt{Im}(\nabla f(\bm{x} + hi \cdot \bm{e}_k))}{h}.
\end{equation}
Here, $[\cdot]_k$ gives $k$-th column of Hessian matrix, and $\bm{e}_k$ is the $k$-th canonical basis. Eq.~\ref{eq:csfd-ad} allows us to construct Hessian only with $M$ perturbations (instead of $M^2$ perturbations using high-order CSFD), and those $M$ perturbations can be conveniently parallelized.  

\begin{figure}
  \centering
  \includegraphics[width=\linewidth]{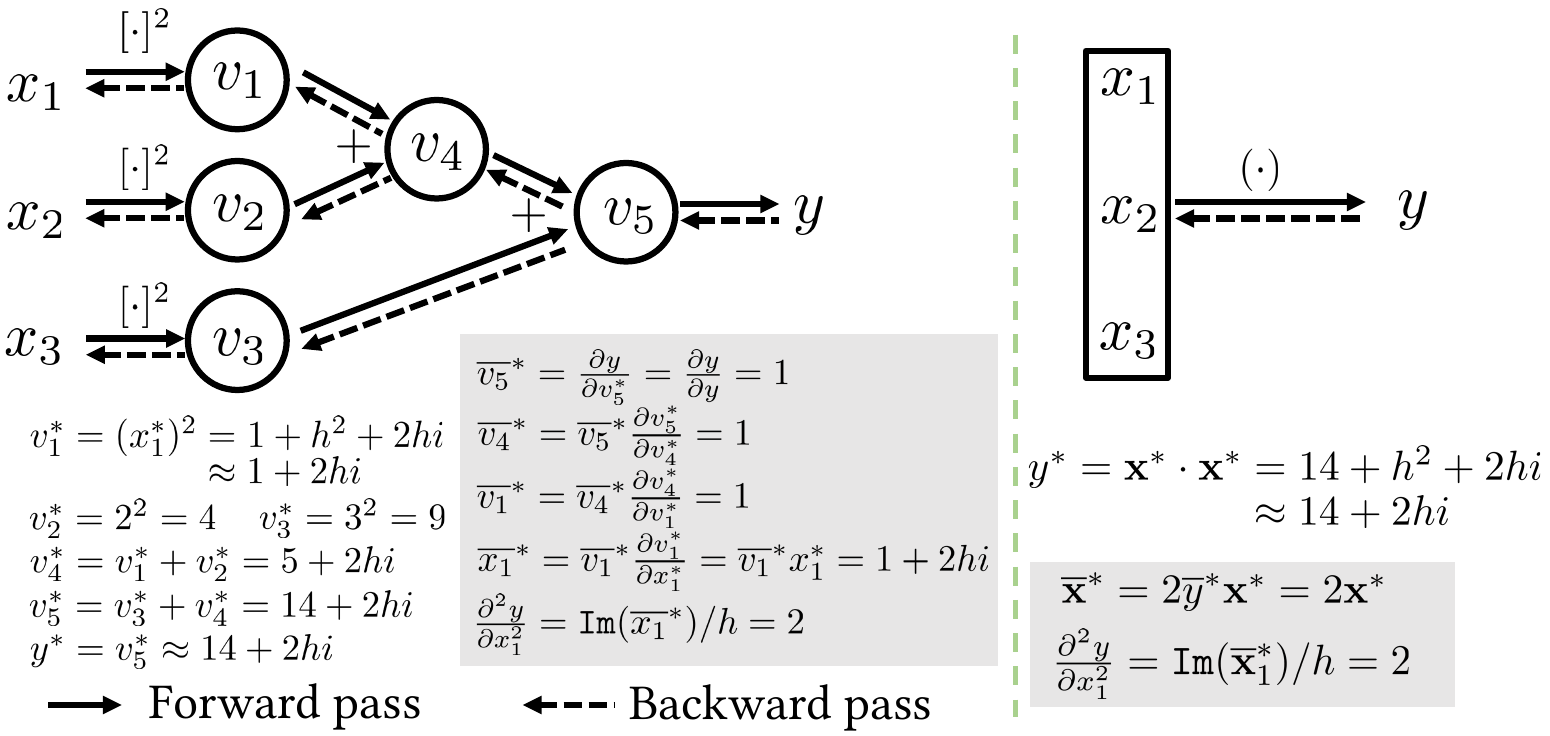}
  \caption{\textbf{CSFD-AD procedure.}~~We show an illustrative example of using CSFD-AD to compute the self inner product of a 3-vector $\bm{x}$: $f=\bm{x} \cdot \bm{x}$.}\label{fig:csfd_ad}
\end{figure}

Fig.~\ref{fig:csfd_ad} shows an illustrative example of CSFD-AD procedure over a simple computational graph of $f=\bm{x} \cdot \bm{x} = x_1^2 + x_2^2+x_3^2$, where $x_1=1$,  $x_2=2$, and $x_3=3$. The forward pass propagates the perturbed $x_1^*=1+hi$ via the intermediate values $v_i^*$ to the final output $y^*=14+2hi$. Note that the high-order term of $h$ in $v_1^*$ is discarded. Following this, the backward pass uses the original AD algorithm but with all adjoints and computations being complex based. The real part of $\overline{\bm{x}}^*$ carries the value of the first-order derivative same as normal AD, while the second-order derivative equals the imaginary part of $\overline{\bm{x}}^*$ upon its division by the input disturbance $h$. We have further optimized such vector and matrix operations. For more details, please refer to Section \ref{sec:matrix} and the right part of this figure.

\subsection{CSFD overload}
Here, we briefly discuss the implementation of CSFD-overload operators and functions in CSFD-AD when used in a computing procedure, as well as their differentiability.

Relational logic operators such as $>$, $\geq$, $<$, $\leq$ and $==$ are common in computing code but usually not defined for complex numbers. They interact with conditional statements to guide execution flow. We make sure that the results of the original algorithm and its CSFD-overload version match each other e.g., $a^* > b^*$ for $a^*, b^* \in\mathbb{C}$ should always yield the same result as $a > b$. This is done by setting $a^* > b^* \leftarrow \mathsf{Re}(a^*) > \mathsf{Re}(b^*)$. Similarly, single-argument functions such as $\max$ or $\min$ only evaluate the real parts of the input.

Conditional statement-driven computations are often discontinuous, leading to multiple branches such as: $f(x) = x$ for $x \geq 0$ and $f(x) = 2x$ for $x < 0$. CSFD applies the perturbation in the imaginary direction, which is orthogonal to the real axis. As a result, it always returns the differentiation of the function at the branch of $f\big(\mathsf{Re}(x_0^*)\big)$. In this example, if $\mathsf{Re}(x_0^*) \geq 0$, CSFD yields the gradient of $f(x) = x$. On the other hand, if $\mathsf{Re}(x_0^*) < 0$, CSFD yields the gradient of $f(x) = 2x$. There is no special treatment of safeguard needed.



Arithmetic operators include $+$, $-$,$\times$, $/$ and functions include $\mathsf{sin}$,  $\mathsf{log}$, $\mathsf{exp}$, just name a few. Most of these elementary functions are \textit{analytic} on complex domain, i.e., the complex function $f^*(a + bi) = u(a, b) + v(a, b)i$ satisfies the Cauchy-Riemann equations:
\begin{equation}
    \frac{\partial u}{\partial a}=\frac{\partial v}{\partial b}, \quad \frac{\partial v}{\partial a}=-\frac{\partial u}{\partial b}.
\end{equation}
A complex function that is analytic exhibits differentiability on its domain around $x^*_0=a_0 + b_0i$ means that it can be differentiated infinitely many times, and CSFD-AD always yields correct results. For instance, the complex promotion of 
$\sin(x)$ is $f^*(x^*)=\sin(a + bi)=\sin(a)\cosh(b) + \cos(a)\sinh(b)i $, since $\frac{\partial u}{\partial a}= \cos(a)\cosh(b) = \frac{\partial v}{\partial b}$, and $\frac{\partial v}{\partial a}=\sin(a)\sinh(b)=-\frac{\partial u}{\partial b}$. Similarly, we have $\cos(a + bi) = \cos(a)\cosh(b) - \sin(a)\sinh(b)i$.

During the backward pass of CSFD-AD, the complex-promoted adjoint is $\overline{x}^* = \cos(x^*) \overline{\sin(x^*)}$, which can be expand as:
\begin{multline}
      \overline{x}^* = \cos(a)\cosh(b)\mathtt{Re}(\overline{\sin(x^*)}) + \sin(a)\sinh(b)\mathsf{Im}(\overline{\sin(x^*)}) \\
    + \left(\cos(a)\cosh(b)\mathsf{Im}(\overline{\sin(x^*)}) - \sin(a)\sinh(b)\mathsf{Re}(\overline{\sin(x^*)})\right)i. 
\end{multline}

In general, if a real-value function $f$ is differentiable around $x_0$, and its complex version is irrelevant to its complex conjugate $x_0^{*^C} = a_0 - b_0i$, it satisfies the Cauchy-Riemann equations and therefore complex-differentiable. 

The absolute value function $\mathsf{abs}$ however, is an expectation. Standard complex $\mathsf{abs}$ function defined as $f^*(x^*)=|a + bi|=\sqrt{x^* \cdot x^{*^C}}=\sqrt{a^2 + b^2}$ always returns a positive non-negative number for complex inputs. It corresponds to the magnitude of $x^*$. While a bit counter-intuitive, this function does not satisfy the Cauchy-Riemann identities, and it is \emph{not} analytic (and cannot be used in CSFD). To this end, we derive an analytic form of the absolute value function:
\begin{equation}
\mathsf{abs}(a + bi)=
\left\{
\begin{array}{r}
-a- bi, \quad \text { if }  a<0, \\
ax + bi, \quad \text { if }  a \geq 0.
\end{array}
\right.
\end{equation}
When CSFD-AD is applied, the adjoints are:
\begin{equation}
\overline{x}^*=
\left\{
\begin{array}{r}
-\overline{\mathsf{abs}(a + bi)}, \quad \text { if }  a < 0, \\
\overline{\mathsf{abs}(a + bi)}, \quad \text { if }  a \geq 0.
\end{array}
\right.
\end{equation}

The chain rule forms the foundation of AD, which allows local differentials to be assembled to be the global differentiation. CSFD-AD also uses the chain rule in the backward pass. 
The generalized chain rule for complex valued function $y^*=f^*(x^*)$ and $z^*=g^*(y^*) = g^*\big(f^*(x^*)\big)$ can be applied as:
\begin{equation}
\frac{\partial z^*}{\partial x^*}=
\frac{\partial g^*(y^*)}{\partial x^*} =\frac{\partial g^*}{\partial y^*} \frac{\partial y^*}{\partial x^*} + \frac{\partial g^*}{\partial y^{*^C}} \frac{\partial y^{*^C}}{\partial x^*},
\end{equation}
If a function is analytic, the second term reduces to zero, and the equation restores to the normal chain rule for real numbers. Under this circumstance, the back-propagation process of CSFD-AD is consistent with AD, except that the computation rule needs to follow the complex number operation.

\subsection{Matrix data structures and arithmetic} \label{sec:matrix}
We use the \texttt{Eigen} \texttt{C++} Library as our primary complex-value matrix and vector container.
Applying AD to some matrix operations could produce a large number of redundant nodes and edges on the computational graph. For example, the squared norm of an $N$-dimension vector is computed as: $ \|\bm{x}\|^2 = \bm{x} \cdot \bm{x} =x_1 \cdot  x_1 + x_2 \cdot x_2 + \cdots + x_N \cdot x_N$. It takes $N$ multiplications and $N-1$ additions, and therefore introduces $2N-1$ nodes for storing and computing intermediate results and adjoints. We eliminate this memory and computing overhead by implementing a specialized function for $y = f(\bm{x}) = \|\bm{x}\|^2$: $\overline{\bm{x}}  = 2 \overline{y} \bm{x}$, which only needs $N$ adjoints for $\overline{\bm{x}}$. 

\subsection{Linear algebra decomposition and factorization}
We implement CSFD-AD to support most known linear algebra functions including trace, determinant, inverse and singular values. We list a few commonly used ones in Tab.~\ref{tab:complexmatrix}. 
{\small \fontfamily{ppl}\selectfont
\begin{table}
\caption{\textbf{CSFD-AD for matrix operation.}~~We list CSFD promotions of matrix functions that we have implemented for elastic locomotion.}\label{tab:complexmatrix}
\begin{center}
\begin{tabular}{l|l||l|l}
\whline{1.15pt}
Function & $\overline{\bm{X}}$ &  Function & $\overline{\bm{X}}^*$ \\
\whline{0.65pt}
$z^*=\|\bm{x}^*\|^2                   $  &  $2 \overline{z}^* \bm{x}^*$  &
$z^*=\bm{x}^*{^\top} \bm{y}^*       $  &  $\overline{z}^* \bm{y}^*$  \\
$\bm{Z}^*=\bm{X}^*\bm{B}^*          $  & $\overline{\bm{Z}}^* \bm{B}^*{^\top}$  &
$\bm{Z}^*=\bm{A}^*\bm{X}^*          $  & $\bm{A}^*{^\top} \overline{\bm{Z}}^*$ \\
$\bm{z}^*=\bm{A}^*{^{-1}}\bm{x}^*   $  & $\bm{A}^*{^{-\top}}\overline{\bm{z}}^*$ &
$\bm{z}^*=\bm{X}^*{^{-1}}\bm{b}^*   $  &  -$\bm{X}^*{^{-\top}}\overline{\bm{z}}^*\bm{z}^*{^\top}$ \\
$z^*=\operatorname {tr} (\bm{X}^*{^{\top}}\bm{X}^*)   $  & $2\overline{z}^*\bm{X}^*$ & 
$z^*=\mathsf{det}({\bm{X}^*})         $  & $\overline{\bm{z}}^*\bm{z}^*\bm{X}^*{^{-\top}}$ \\
\whline{1.15pt}
\end{tabular}
\end{center}
\end{table}
}
Singular value decomposition (SVD), $\bm{A}=\bm{U}\bm{\Sigma}\bm{V}^\top$, is a widely used numerical procedure in FEM simulation e.g., to extract the invariants of the deformation gradient. While SVD of such $3 \times 3$ has a closed-form formulation, resulting singular values of a complex matrix are always real (e.g., similar to complex $\mathsf{abs}$). Therefore, SVD operation is not analytic, and CSFD promotion cannot be applied to its closed form. Alternatively, we apply CSFD-AD to the iterative numerical SVD procedure for real matrices and overload each underlying operation with complex arithmetics. In our implementation, we employ implicit-shifted symmetric QR SVD~\cite{gast2016implicit} for fast and robust decomposition on symmetric $3 \times 3$ matrices.  


Elastic locomotion needs to solve a sparse system at each Newton iteration in the form of $\bm{A}\bm{x} = \bm{b}$. We use direct solvers like \texttt{LU}, \texttt{LLT}, and \texttt{LDLT} to handle those linearized system solves. Unlike SVD, matrix factorization is analytic, and CSFD promotion can be directly applied. However, doing so expands the computational graph exponentially in the first AD pass. Therefore, we compute the adjoints of the solution of the linear solve: $\bm{b}=\bm{A}^{-1}\bm{x}$. 
During the forward pass, the factorized matrix of $\bm{A}$ is stored; and in the backward pass, the adjoint of the linear solve i.e., $\overline{\bm{b}}$ is then obtained by solving $\overline{\bm{b}}=\bm{A}^{-\top}\overline{\bm{x}}$ using the saved factorization of $\bm{A}$. This strategy is similar to the adjoint method~\cite{givoli2021tutorial}, which frees us from obtaining the actual adjoints of the matrix inverse. Alternatively, it is also possible to apply CSFD promotion to a numerical iterative linear solver as in many existing differentiable simulation frameworks~\cite{hu2019difftaichi,du2021diffpd}.  Doing so requires the total number of iterations to be pre-known. Therefore, existing methods often use a prescribed iteration count. This is a risky measure: too few iterations will cause the solver to diverge, while an over-conservative setting consumes too much memory since the computational graph expands proportional to the number of iterations. Elastic locomotion involves nonlinear contact barriers, and the variation of iteration counts is substantial depending on the actual configuration of contacts and frictions. Therefore, assuming a fixed iteration number is not an option for us.

\section{Locomotion Control}\label{sec:control}
The mixed differentiation modality with CSFD-AD offers a generic way to evaluate high-order differentiation of a complex computation procedure. Therefore, our system is compatible with a wide range of intuitive controllers to allow the user to manipulate, author, and design the locomotion of soft objects via high-level kinematic objectives. We also design an intuitive user interface to facilitate such edits. In this section, we describe some loss functions used in our examples. 

\subsection{Target functions}
\paragraph{Linear motion controllers.}
We achieve the desired locomotion mainly by controlling the target trajectory of some key points on the object. For the position of a target point $\bm{x}^\star$, the objective function that constrains the position can be written as:
\begin{equation}
    {G}_{position} = \left\|f(\bm{x}) - \bm{x}^\star \right\|^2, 
\end{equation}
where $f$ could refer to the calculation of the COM (center of mass) $\bm{c}=f_{COM}(\bm{x})$ of the body, of local COM coordinates (foot, head, base, etc.), or of the relative position between points. We also use similar objective functions to track the velocity: 
\begin{equation}
    {G}_{vel.} = \left\|f\left(\frac{\bm{x} - \bm{x}_t}{\Delta t}\right) - \bm{v}^\star \right\|^2, 
\end{equation}
and acceleration:
\begin{equation}
    {G}_{acc.} = \left\|f\left(\frac{\bm{x} - 2\bm{x}_t + \bm{x}_{t-1}}{\Delta t^2}\right) - \bm{a}^\star \right\|^2.
\end{equation}
For instance, we could specify the starting speed of the character to achieve jumps of different heights in addition to its COM target, and we control the relative speed of the character's head and base to achieve the aerial posture of the forward jump. We can achieve object lift-off by setting the COM acceleration $\bm{a}^\star$ equal to the gravitational acceleration. The control over the COM acceleration is also equivalent to the control of the derivative of the linear momentum $\dot{\bm{L}}=m\ddot{\bm{c}}$, which takes part in balance control~\cite{tan2012soft}.

\paragraph{Angular motion controllers.}
We control the rotation of the body around an axis by manipulating the changes in the body's angular momentum:
\begin{equation}
    {G}_{angular} = \left\| \dot{\bm{L}}_{angular}(\bm{x})- \dot{\bm{L}}^\star \right\|^2, 
\end{equation}
where $\bm{L}_{angular}=\sum_i m\bm{r}_i \times \bm{v}_i$ is the total angular momentum of the collection of points. This objective is typically used to control the rotation and twisting of an object on the ground, such as allowing the character to gain sufficient angular velocity before it flips and jumps in the air. If a body that wants to perform a flip movement in the air, we control the body's moment of inertia (MOI) $\bm{I} = \sum_i m_i \bm{r}_i^2$ around the axis to make the character contract and rotate faster. This is also a common technique used by diving athletes.
\begin{equation}
    {G}_{MOI} = \left \| \dot{\bm{I}}(\bm{x}) - \dot{\bm{I}}^\star \right\|^2.
\end{equation}

\paragraph{Additional control variants.}
Thanks to the CSFD-AD differentiation algorithm, our framework allows users to use sophisticated target functions without manually deriving their closed-form derivatives. For instance, controlling the changes in elastic energy of an object:
\begin{equation}\label{eq:g_energy}
{G}_{elastic} = \left\|\int \dot{\Psi} \mathrm{d}V - \dot{E}^\star\right\|^2,
\end{equation}
where the integral applies over a specific portion of the body. This objective function plays an essential role in the jumping experiment shown in Fig.~\ref{fig:demo_trampoline}. By specifying the change in the potential energy of the trampoline, we can control the future kinetic energy of the lamp, allowing the lamp to reach different heights.

We also control the size of the projection base area, allowing the object to expand its base to maintain balance during the landing process of continuous jumps:
\begin{equation}
{G}_{proj} = \left\| \dot{A}_{proj}(\bm{x}) - \dot{A}^\star \right\|^2,
\end{equation}
where $A_{proj}$ returns the projection area of the body on some contacting or supporting plane. The strategy of designing and choosing the most effective controller is the core problem of animation control and a wide range of specific motion planning problems. This is not the focus of this work. Nevertheless, we do believe CSFD-AD delivers a useful computational tool, and it could enable more creative and versatile controllers. 


\subsection{Regularization terms}\label{sec:regularization}
We design regularization terms to enhance the stability of the optimization process, i.e., the Hessian of the objective function Eq.~\eqref{eq:loss} is positive definite. This is similar to the LM method~\cite{more2006levenberg}. Specifically, we regulate the total muscle energy to avoid conflicting or excessive muscle actions that would cause energy waste and trigger improbable distortion:
\begin{equation}
    R_{energy} = \frac{1}{2}k\|\bm{a}\|^2, 
\end{equation}
where $k > 0$ controls the strength of the regularization.

In addition, we impose penalties on muscle segments whose rate of change $|\dot{a}_i|$, exceeds a pre-defined threshold $\dot{a}_{max}$:
\begin{equation}
R_{change} = \sum_{i} \big( |\dot{a}_i| - \dot{a}_{max} \big)^2, \quad \forall |\dot{a}_i| > \dot{a}_{max}.
\end{equation}


\begin{figure*}
    \centering
    \includegraphics[width=\linewidth]{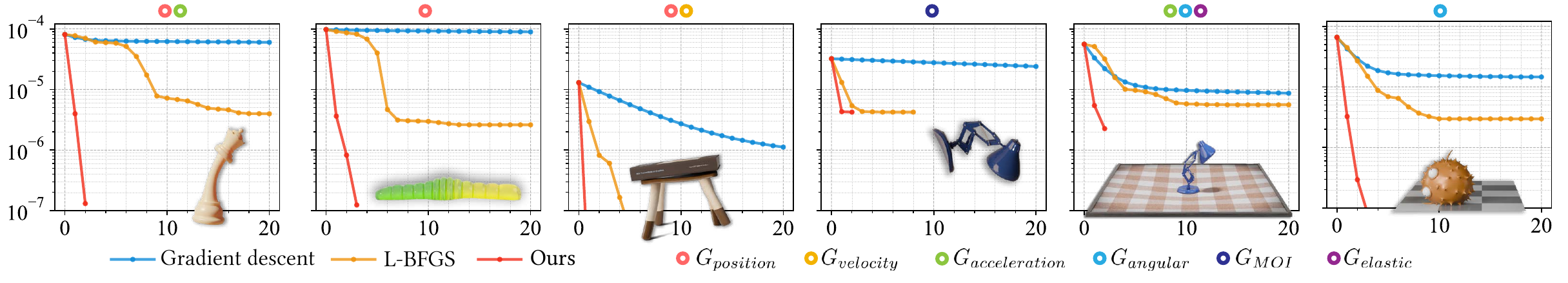}
    \caption{\textbf{Convergence curves under different combinations of losses.}~~We compare the convergence performance using gradient descent~\cite{hu2019difftaichi}, quasi-Newton (L-BFGS)~\cite{du2021diffpd,huang2022differentiable}, and full Newton (our method) at some representative frames of different elastic locomotions and combinations of loss functions. Newton's method consistently outperforms other optimization algorithms after a proper warm start. If the body does not undertake extensive collisions (e.g., the third and fourth plots), L-BFGS also yields good convergence.}
    \label{fig:vsLBFGS}
\end{figure*}

{\small \fontfamily{ppl}\selectfont
\begin{table}
\caption{\textbf{Time statistics.}~~This table reports simulation setups and time performance of examples shown in the paper. $\#$ Ele. is the total numbers of elements on the body. $|\bm{a}|$ is the dimension of control variables i.e., muscle activations. $N_c$ reports the maximum number of contacts processed per time step. Sim, Grad, Hess and Solve give the average timing of forward simulation, gradient computing, Hessian computing and total solving time in seconds per time step.}\label{table:time}
\centering
\begin{tabular}{l|c|c|c|c|c|c|c}
\whline{1.15pt}
Soft body & $\#$ Ele. & $|\bm{a}|$  & $N_c$  &  Sim  & Grad & Hess & Solve\\
\whline{0.65pt}
Caterpillar {\scriptsize Figs.~\ref{fig:vsLocomotion},\ref{fig:worm}} & 2.6K & 62   & 140         & 0.8  & 0.5  & 5.2       &   15.4   \\
Starfish {\scriptsize Fig.~\ref{fig:demo_starfish}} & 2.8K & 66   & 192         & 0.9  & 0.5  & 6.2       &   20.3   \\
Chess {\scriptsize Fig.~\ref{fig:demo_chessJump}} & 8.4K & 72    & 224         & 1.4  & 1.0  & 12.8    &   36.6   \\
Lamp {\scriptsize Figs.~\ref{fig:teaser}, \ref{fig:demo_trampoline}}       & 5.2K & 68    & 166        & 1.1   & 0.9    & 8.8     &   19.5   \\
Stool {\scriptsize Fig.~\ref{fig:demo_stoolWalk}} & 6.8K & 40    & 246          & 1.0  & 0.6  & 4.8      &   21.6   \\
``T'' {\scriptsize Fig.~\ref{fig:demo_T_rotate}} & 1.2K & 51    & 25          & 0.2  & 0.2  & 3.4      &   5.5  \\
Puffer fish {\scriptsize Fig.~\ref{fig:demo_pufferfish}} & 27.6K &  32  & 171        & 4.7  & 2.8    & 27.1  &   68.6     \\
\whline{1.15pt}
\end{tabular}
\end{table}
}

\section{Result}\label{sec:result}

We implemented our framework on a desktop computer with an \texttt{Intel} \texttt{i9-13900KF} CPU with 128GB of memory. We developed our CSFD-AD framework based on~\texttt{Stan Math} library~\cite{carpenterStanMathLibrary2015} (for reverse AD) and \texttt{Eigen} Library ~\cite{eigenweb} (for complex linear algebra and matrix interface). We implemented the complex promotion of all analytic functions by ourselves. We enabled Intel \texttt{TBB} ~\cite{pheattIntelThreadingBuilding2008} for parallel processing of Hessian calculations in multiple threads. The detailed statistics of experiments are reported in Table~\ref{table:time}. The time step size is set as $\Delta t = 1/40$ sec in all the experiments. Please refer to the supplementary video for animated results.


\subsection{Comparison with gradient descent and L-BFGS}
\begin{wrapfigure}{r}{0.5\linewidth}
    \includegraphics[width=\linewidth]{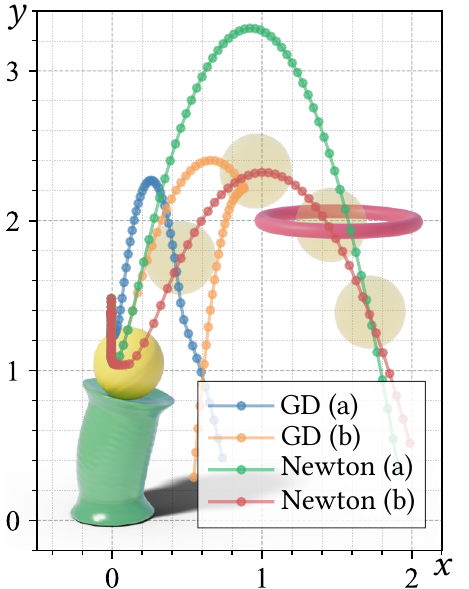}
    \caption{\textbf{Comparison with gradient descent.}~~An I-shaped soft-body player tries to ``shoot'' a ball into a basket under user-specific trajectory. Gradient descent fails to converge the loss sufficiently, while the soft-body player always secures the score with Newton's method (ours).}\label{fig:vsgd}
\end{wrapfigure}

The first experiment we would like to report is to highlight the need of a second-order optimizer of inverse control problems. To this end, we compare the results using gradient-based methods and Newton's method through a basketball shooting experiment, as shown in Fig.~\ref{fig:vsgd}. In this experiment, an I-shaped soft body pushes a ball to pass a basket ring. Observing the projection plane, the initial position of the ball is $(0, 1.5)$, and the basket of radius $0.5$ is at $(1.5, 2)$. The body first contracts, bends forward and then releases. Before releasing, it has to ensure that the relationship between the speed and position of the ball's COM is satisfied in order to generate the right contact force to push the ball to fall into the basket. 

In the figure, the blue curve GD (a) visualizes the trajectory if we only use gradient information (with $1,000$ iterations) for solving Eq.~\eqref{eq:loss} e.g., as in~\cite{hu2019chainqueen,hu2019difftaichi}. The ball fails to follow the prescribed path and does not even reach the basket due to lesser convergence. The green curve Newton (a) is the result of our method with second-order optimization -- the ball passes through the basket ring accurately. Even if we switch to gradient descent only at the last time step before the body releases the ball, the resulting trajectory is still inaccurate (the orange curve GD (b)), where the ball bounces back from the rim. The red curve Newton (b) shows the result of second-order optimization for another target trajectory, where the soft-body player scores again.

\begin{figure}
    \centering
    \includegraphics[width=\linewidth]{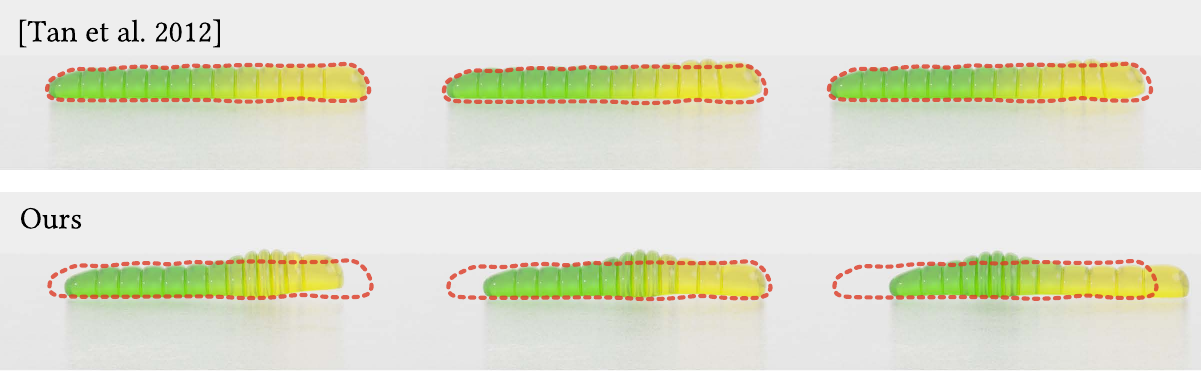}
    \caption{\textbf{Gummy caterpillar (comparison with~\cite{tan2012soft}}~~A soft gummy caterpillar moves on the ground. Such motion is enabled by a combination of stretching and contracting of the body under persistent body-ground contacts. Such sophisticated locomotion involves a large number of contact points and frequent switches between sticking and sliding friction. LCP-based solvers~\cite{tan2012soft} become extremely slow in finding a usable solution, and the caterpillar barely moves. Our method, on the other hand, leveraging IPC and line-search Newton, helps the caterpillar move forward as expected.}
    \label{fig:vsLocomotion}
\end{figure}

We further compare our method with gradient descent and L-BFGS~\cite{huang2022differentiable,du2021diffpd} in several representative scenes of different combinations of loss functions. The results are shown in Fig.~\ref{fig:vsLBFGS}. In the figure, $x$-axis represents the number of iterations after the same warm start, and $y$-axis is the value of the loss function (the low the better). Our method demonstrates strong second-order convergence in all examples. The convergence gap is widened if the specific frame contains a large number of frictional contacts. On the other hand, there is no contact, L-BFGS also gives a good convergence (e.g., see the fourth plot in Fig.~\ref{fig:vsLBFGS}). Meanwhile, gradient descent does not offer the desired convergence in all cases.


\subsection{Comparison with LCP-based contact resolution}
The contact and friction between the soft body and the environment are often the most challenging aspects of locomotion control. Complementarity programming is an accurate math model of the contract problem but its convergence is not guaranteed. This drawback becomes more severe for elastic locomotion because a soft body often uses wide contact regions to adjust its pose and motion. For instance, \citet{tan2012soft} solve the body collision using the QPCC solver. In~\cite{tan2012soft}, each contacting point has 10 contact states, and the minimizer is found in the solution space of $10^{N}$. In the worst case, they need to perform an exhaustive search to obtain a global minimizer. This is prohibitive for soft bodies of moderately high resolution, as we can easily have dozens of contact points. An early termination of the combinational search could leave the body in an ill-defined configuration. To mitigate this difficulty, \citet{tan2012soft} only supports the solution of four contact patches. This, in turn, reduces the accuracy of the contact model. 

Our framework follows the interior-point method to use impulse-like (but smooth) contact barriers to model contact and friction. As a result, our method supports complex contact modeling. A representative example is shown in Fig.~\ref{fig:vsLocomotion}. In this example, we control a gummy caterpillar to crawl forward on the ground. There are 140 contact points in total, and the muscles embedded trigger alternations between contractions and expansions at body parts contacting the ground. The contact states of those points vary at each frame. In this case, the QPCC solver and search strategy proposed in~\cite{tan2012soft} cannot obtain a usable approximate within a reasonable amount of time. Our method, on the other hand, is less sensitive to the number of contact points. The strong convergence of Newton's method robustly helps find a good solution at each frame.

\begin{figure*}
  \centering
  \includegraphics[width=\textwidth]{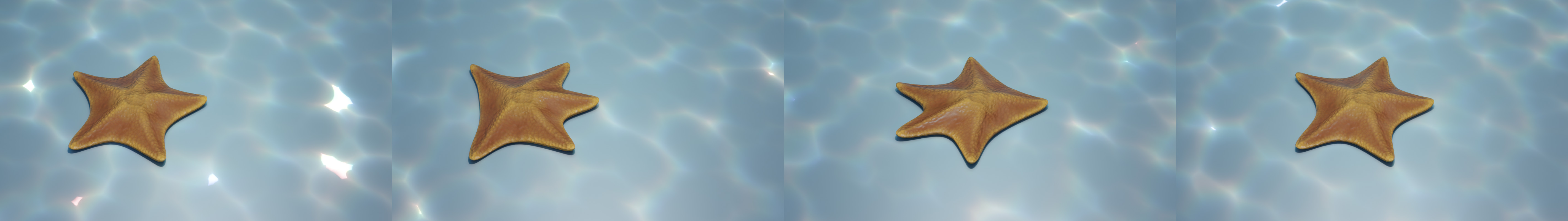}
  \caption{\textbf{Crawling starfish.}~~A starfish moves its legs, adjusts its posture, and displaces to the right. This locomotion involves 192 contact points. LCP-based methods are unable to find a good solution. }\label{fig:demo_starfish}
\end{figure*}
\begin{figure*}
  \centering
  \includegraphics[width=\textwidth]{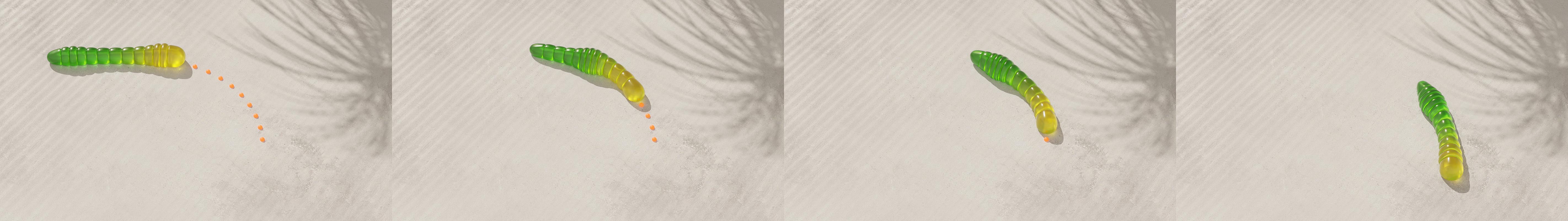}
  \caption{\textbf{Crawling caterpillar.}~~A gummy caterpillar (e.g., the same model in Fig.~\ref{fig:vsLocomotion} crawls following the user-provided trajectory. This motion is more challenging than the one shown in Fig.~\ref{fig:vsLocomotion} since the friction is asymmetrical along the body in order to generate the curved trajectory.}\label{fig:worm}
\end{figure*}

\subsection{More results}
We have tested elastic locomotion under an array of scenarios. Thanks to the accurate and efficient evaluation of the Hessian, our method allows the user to animate elastic bodies in various forms and produce interesting animations.

\paragraph{Crawling.} In nature, soft-bodied creatures such as worms, slugs, caterpillars, and certain types of mollusks and aquatic animals exhibit unique methods for crawling. Fig.~\ref{fig:demo_starfish} shows a crawling animation of a starfish. 

One common crawling method seen in soft creatures is called peristalsis or wave-like motion. For instance, earthworms move through soil using this principle. Their bodies, made of a long series of muscle segments, alternately contract and expand, generating a wave from the anterior to the posterior of the body. As each segment of the worm contacts the soil, it expands, generating friction against the ground. This wave-like motion propels the animal forward through the environment. 

We demonstrate such locomotion using a model of a gummy caterpillar in Fig.~\ref{fig:worm}. The target involves dividing the gummy worm model into five separate segments from head to tail to detail the varying states of contraction and relaxation. It implements a sinusoidal function to generate the trajectory of these five parts due to their inherent periodicity, which mirrors the cyclical pattern of muscle movement. Inside the body of the model, we have placed four longitudinal muscles and seven circular muscles. The optimized result is realistic and interesting. Beginning with the longitudinal muscles, their contractions shorten the body, causing it to broaden and rise slightly. Once these muscles relax following the contraction, the body then elongates and flattens out, reaching further forward. On the other hand, the circular muscles run around the body vertically. When these muscles contract, they cause the body to elongate and thin out. Afterward, when the circular muscles relax, the body returns to its original, more rounded shape. 

\begin{figure*}
  \centering
  \includegraphics[width=\textwidth]{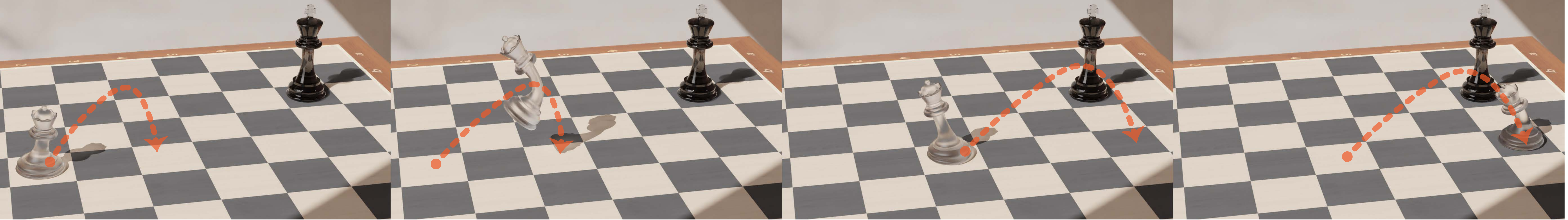}
  \caption{\textbf{Jumping chess.}~~The queen in chess performs consecutive leaps forward to declare a check. This example also shows that we may avoid a full space-time optimization if the inverse problem at each frame can be accurately solved.}\label{fig:demo_chessJump}
\end{figure*}
\begin{figure*}
  \centering
  \includegraphics[width=\textwidth]{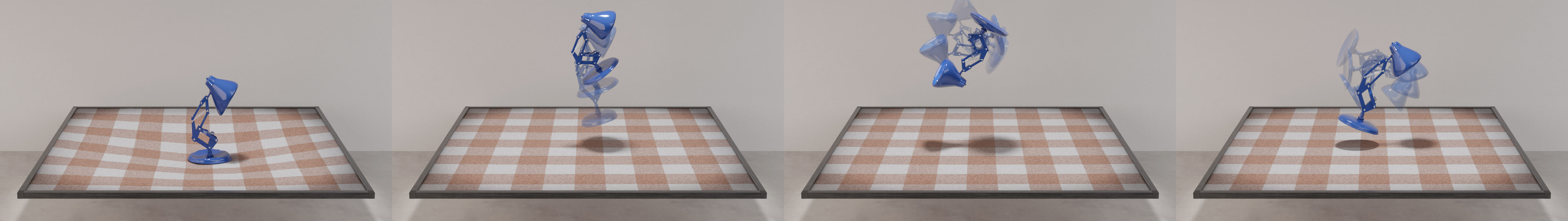}
  \caption{\textbf{Trampoline acrobatics.}~~Our method is able to handle high-dimension and complex contact between the body and deformable objects. In this example, a lamp jumps on a trampoline doing a backflip. We adjust the target elasticity potential Eq.~\eqref{eq:g_energy} to control the level height of each jump.}\label{fig:demo_trampoline}
\end{figure*}
\paragraph{Jumping.} Jumping soft-bodied characters are more commonly seen in cartoons and animations. 
We have replicated the jumping action of Pixar's Luxo Jr., but in a soft-bodied and physics-based manner. In addition, we have also designed an animation of a game of chess, where the soft-bodied chess pieces move across the board by continuously jumping, as shown in Fig.~\ref{fig:demo_chessJump}.

We divide the jumping process into takeoff phase, mid-air phase and landing phase and apply different controllers accordingly. During the takeoff phase, we first control the position and velocity of COM to compress the body then quickly reverse into a rapid relaxation, releasing the built-up potential energy and increasing contact against the ground to overcome gravity force and follow the target trajectory. The jumping direction and velocity are more accurately handled by our second-order optimization process since the soft body doesn't have control over its trajectory during the mid-air phase. During the mid-air phase, we control the relative velocity of the center of base (COB) and COM to make a fore jump or the angular velocity to make a flip. After that, the soft body expands its base and decreases the velocity of COB to make a softer landing. The body can then immediately start preparing for another jump, or continue with other types of locomotion.

In our teaser figure (Fig.~\ref{fig:teaser}), the lamp jumps onto the stool and then onto the glass coffee table. The friction coefficient of the glass is smaller (0.1), so the lamp controls the position of COM and the contact surface to slide on the table without losing balance. Then, it performs a front flip. During the flip, the lamp reduces its moment of inertia to obtain a greater angular speed. After landing, the lamp rotates and jumps towards the camera.

We also demonstrate a motion involving the collision between the soft character and the deformable environment. In Fig.~\ref{fig:demo_trampoline}, the lamp bounces on a deformable trampoline multiple times. When it lands on the trampoline, the objective function is designed as the total elastic potential energy of the trampoline. In the first three jumps, we increment the elastic potential energy objectives so that the lamp jumps higher each time. During the mid-air phase. We switch between different objective functions to make the lamp maintain its shape, backflip and rotate.

\begin{figure*}
  \centering
  \includegraphics[width=\textwidth]{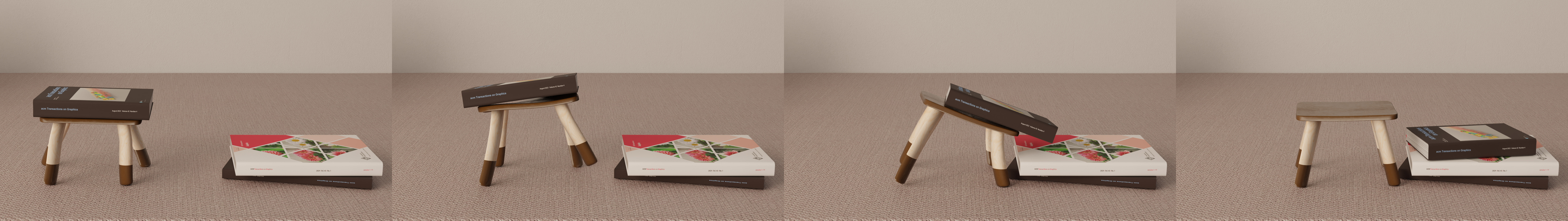}
  \caption{\textbf{A librarian stool.}~A soft robot stool carries a book, walks like a quadruped and leaves the book to slide down and stack on another book. This animation can be intuitively generated by combining several kinematic targets. Our method resolves the stool-ground contacts and stool-book contacts to allow the stool to complete this mission successfully. }\label{fig:demo_stoolWalk}
\end{figure*}

\paragraph{Walking.} 
Fig.~\ref{fig:demo_stoolWalk} shows a walking example. We create the walking pattern by lifting and moving one leg from one side while simultaneously doing the same with the opposite diagonal leg. The stool carries a heavy book, moves forward carefully and lets the book slide down slowly. The target functions involve trajectories of four feet to move forward, as well as the position and direction of the carried book to control it to being held when moving or slid down to other books.

\begin{figure*}
  \centering
  \includegraphics[width=\textwidth]{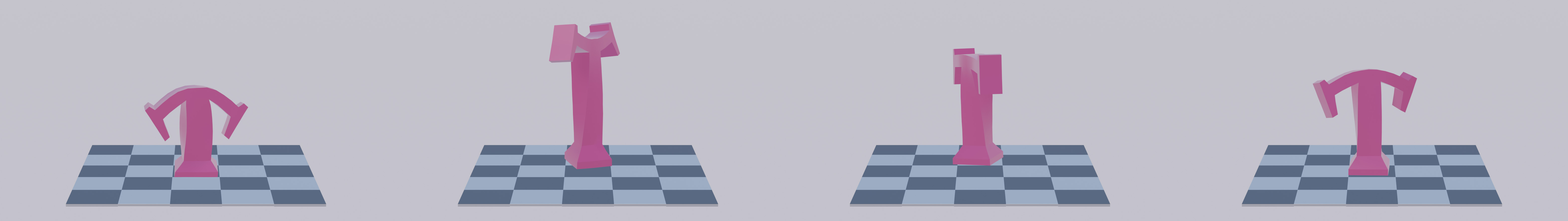}
  \caption{\textbf{T jumps and rotates.}~~This is a classic locomotion example. A T-shaped body jumps, lands, and rotates. We could replicate the locomotion result of~\cite{tan2012soft}.}\label{fig:demo_T_rotate}
\end{figure*}
\begin{figure*}
  \centering
  \includegraphics[width=\textwidth]{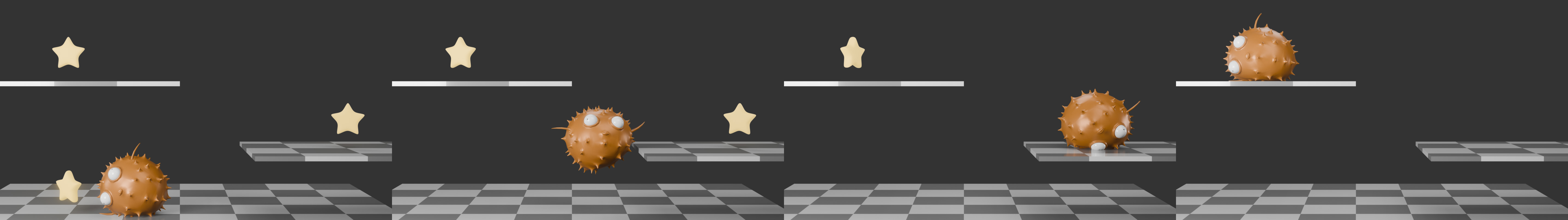}
  \caption{\textbf{Puffer fish's playground}. The spiky puffer fish rolls and leaps between three planks, taking the stars at the target location.}\label{fig:demo_pufferfish}
\end{figure*}

\paragraph{Rolling \& rotating.} 
We first show a classic locomotion result of a T-shaped character. In this example, the ground contact is relatively simple. We find that both our method and \cite{tan2012soft} can produce similar results following the user's input. The T-body jumps, lands, and rotates just like a live creature.

A more interesting and complicated scene is shown in Fig.~\ref{fig:demo_pufferfish}. A puffer fish rolls left and right, jumps onto the platform, and reaches the star at the target location. We have added four ring muscles inside the puffer fish to allow for local deformations and used $G_{angular}$ to control its angular momentum changes. We observed that the puffer fish can move its center of mass in front of the center of pressure to continually roll forward. Thanks to the robustness of IPC and Newton, the spikes of the puffer fish can be reasonably deformed during movement without incurring inversion or collapse. Unfortunately, \cite{tan2012soft} is unable to generate this motion with QPCC.


\section{Conclusion \& limitation}\label{sec:conclusion}
This paper shows a novel differentiation strategy combining numerical differentiation (CSFD) and analytic differentiation (reverse AD) to enable efficient calculation of Hessian of a barrier-in-the-loop inverse simulation procedure. We apply this method to calculate the optimal muscle activations, which drive the locomotion of soft bodies under frictional contacts. Our approach is general, and we anticipate it will benefit other application areas where stability is at odds with constrained optimization. Our choice to focus on soft-body control aims to address a lack of solutions for employing such characters in animation, but the approach we introduce has no direct ties to the specifics of deformation simulation and control, such as robotics and fabrication.

Our method also has limitations. While CSFD-AD is more efficient in calculating the Hessian automatically than existing differentiation schemes, assembly of a large-scale Hessian is expensive and at least $\bm{O}(N^2)$. It is possible to integrate CSFD-AD with other optimization algorithms that do not need the Hessian explicitly, such as L-BFGS. We choose reverse AD for gradient calculation because our loss function is a many-to-one map. Developing CSFD-AD with forward AD is preferred for computations generating high-dimension output from mapping low-dimension input.

\bibliographystyle{ACM-Reference-Format}
\bibliography{ref}

\end{document}